\documentclass[a4paper,11pt]{article}
\pdfoutput=1 

\usepackage{jinstpub} 
\usepackage{hepparticles}
\usepackage{listings}
\usepackage[dvipsnames]{xcolor}
\definecolor{dkgreen}{rgb}{0,0.6,0}
\definecolor{gray}{rgb}{0.5,0.5,0.5}
\definecolor{mauve}{rgb}{0.58,0,0.82}
\usepackage{lineno}

\def\pdf{probability density function}
\title{\boldmath \texttt{GooStats}: A GPU-based framework for multi-variate analysis in particle physics}

\author[a,b]{X. F. Ding}

\affiliation[a]{Gran Sasso Science Institute, \\67100 L'Aquila, Italy}
\affiliation[b]{INFN Laboratori Nazionali del Gran Sasso, \\67010 Assergi (AQ), Italy}

\emailAdd{xuefeng.ding@gssi.it}

\abstract{\texttt{GooStats} is a software framework that provides a flexible environment and common tools to implement multi-variate statistical analysis. The framework is built upon the \texttt{CERN ROOT}, \texttt{MINUIT} and \texttt{GooFit} packages. Running a multi-variate analysis in parallel on graphics processing units yields a huge boost in performance and opens new possibilities. The design and benchmark of \texttt{GooStats} are presented in this article along with illustration of its application to statistical problems.
}


\begin{document}
\maketitle
\flushbottom

\section{Introduction}
\label{sec:intro}

Enormous execution time has become a challenge for complex statistical analysis in experimental particle physics. Fitting is a statistical analysis technique to optimize unknown parameters in an assumed model aiming at minimizing the discrepancy between the model and data using certain test statistics, such as \(\chi^2\) or the likelihood function. The fitting procedure can be extremely slow for complex models, such as those containing integrals, or those with many unknown parameters. The fitting time can also be very long when the number of bins or the number of unbinned data points is very large. 

This challenge can be overcome if the analysis algorithms and softwares are tuned to allow concurrent execution to be adapted to modern computing architectures. In 2005, Herb Sutter pointed out in his "The Free Lunch Is Over"\cite{Herb2005} that while the number of transistors on CPUs is still increasing, mainstream computers are being permanently transformed into heterogeneous supercomputer clusters. The computing power is improved by increasing the number of computing units rather than by increasing the clock rate of single unit. For example, on one hand, the clock rate of Intel Pentium 4 CPU, released in 2003, is 3.8 GHz\cite{Intel2003}, while the clock rate of recent CPUs, such as Intel Xeon E5-2630 v3 released in 2014, has decreased to 2.4 GHz\cite{Intel2014}; on the other hand, the number of computing units per CPU has increased from 2 (Pentium 4) to 8 (Xeon E5-2630 v3). 
What's more, General Purpose Graphics Processing Unit (GPGPU)\cite{MarkHarris2004}, emerged at the beginning of the twenty-first century, has hundreds of computing units and can launch thousands of computation tasks simultaneously\cite{NVidiaInc.2012}.	

Parallel computing techniques have been widely applied in high energy physics and astrophysics\cite{Gorbunov2011,Chan2013,Ariga2014}. Thanks to the open source project \texttt{GooFit}\cite{Access2014}, the fitting task can also benefit from the parallelization acceleration\cite{Sun2017a,Hasse2017}. However, beyond minimization \texttt{GooFit} does not provide a framework holding services and algorithms performing statistical analysis and manipulating input-output, while these tasks can be quite complex, such as varying fit configurations to understand their influences on physics parameters or fitting multiple datasets simultaneously. A framework that addresses these needs will be handy and save much software development time. These considerations gave birth to \texttt{GooStats}\cite{Ding2018}. \texttt{GooStats} is an open source software. It is interface-oriented and thus flexible and easy to extend. The framework is built upon the \texttt{CERN ROOT}, \texttt{MINUIT} and \texttt{GooFit} packages. Users provide inputs in the format of plain text or \texttt{CERN ROOT} histograms, and this framework will produce outputs in the Portable Document Format (PDF), \texttt{CERN ROOT} \texttt{TF1} and \texttt{TTree} etc. This software has been applied in Borexino Phase-II analysis\cite{TheBorexinoCollaboration2018,Agostini2017,Agostini2017a}. In this paper, we present the design and benchmark of the software, and examples of its application.

\section{Software design}
\label{sec:design}
\texttt{GooStats} is a statistical analysis framework. It is initiated to facilitate the use of computation power of modern GPGPUs for fitting and statistical analysis. It serves as the middle layer between statistical analysis module written by users and the minimization engine \texttt{GooFit}. It provides a complete suit of analysis tools, supports common statistical analysis and also includes algorithms used in Borexino analysis\cite{Agostini2017,Agostini2017a}, such as analytical detector response models and multi-observable likelihoods.

\subsection{Structure of the software and responsibilities of classes}
\begin{figure}[htbp]
\centering 
\includegraphics[width=0.9\textwidth]{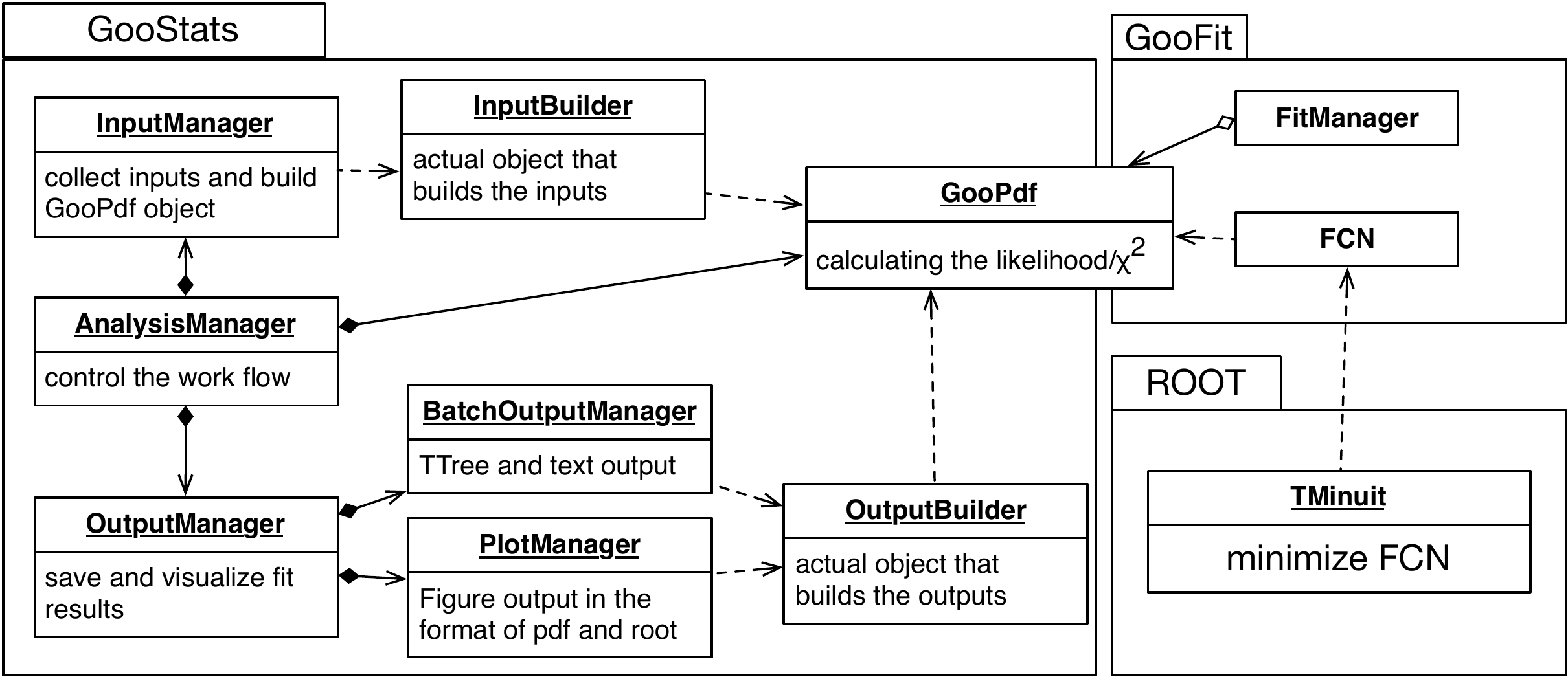}
\caption{\label{fig:GooStats-UML} Simplified structure of \texttt{GooStats} and its relationship with \texttt{GooFit} and \texttt{Minuit}.}
\end{figure}

The structure of \texttt{GooStats} is depicted with a simplified Unified Modeling Language (UML) graph\cite{ObjectManagementGroup2017} in Figure~\ref{fig:GooStats-UML}. 
The classes are named after their responsibilities. The workflow is managed by a singleton of \texttt{AnalysisManager}. Two manager classes are designed responsible for manipulating inputs and outputs, respectively, and the actual work is performed by two builder objects to decouple the realization details. The input-managing object collects and parses the raw data and configurations, and builds with the help of its builder the probability density function objects, which are needed by the minimization engine. The output-managing object collects fit results and produces outputs with the help of two manager objects, responsible for graphic and non-graphic outputs, respectively, and the output builder. The default behavior can be overwritten through the polymorphism mechanism of the \texttt{C++} language. 

\begin{figure}[htbp]
\centering 
\includegraphics[width=0.9\textwidth]{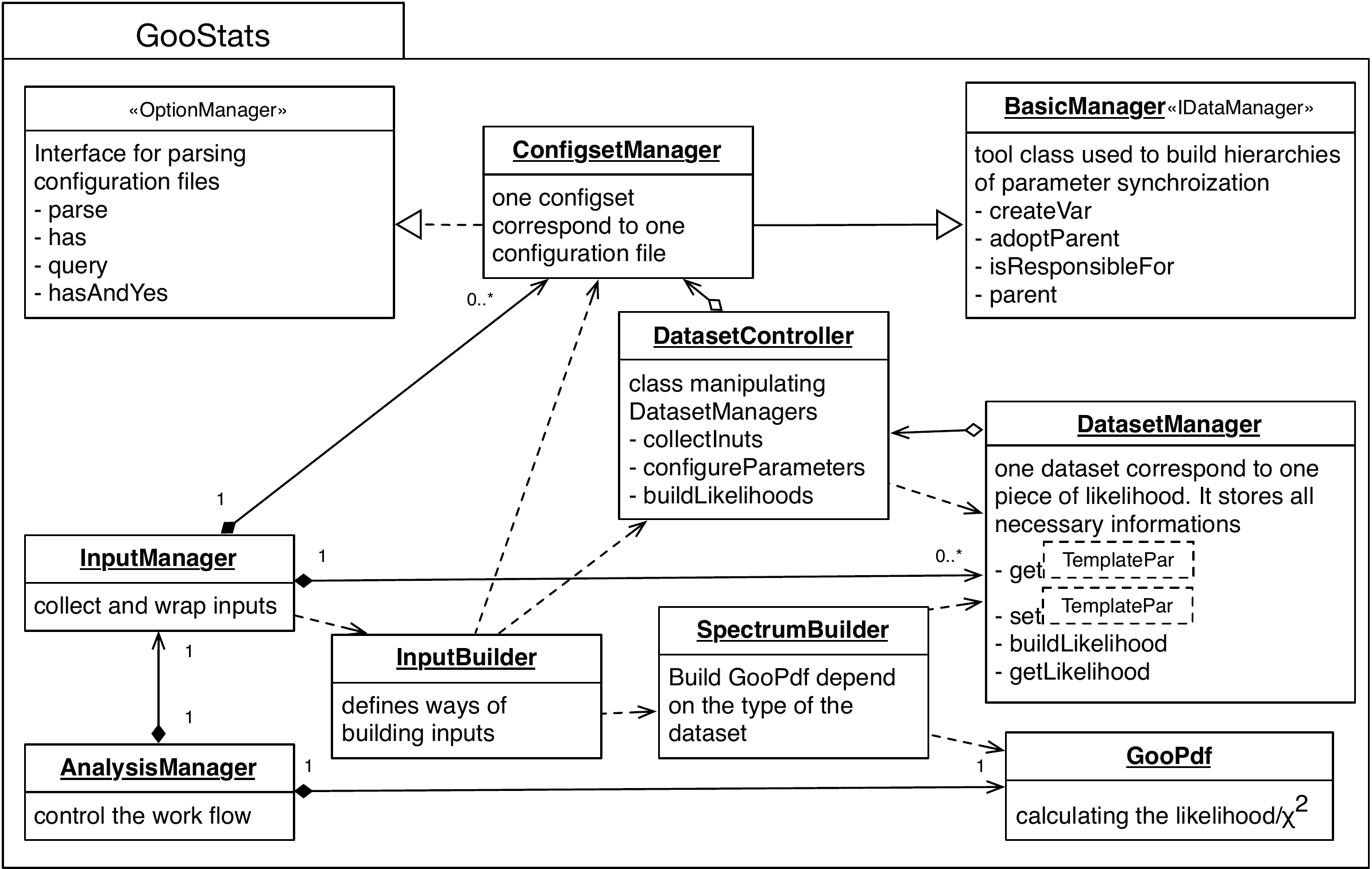}
\caption{\label{fig:GooStats-input-UML} Data structure of \texttt{GooStats} and classes responsible for filling the input}
\end{figure}

The data model is designed to support fit of multiple datasets with independent configurations natively. The information are stored in two levels: configuration associated, which stores parsed key--value pairs, and dataset associated, which stores all essential information to construct a probability density function object including the dataset object, the configurations and the fit parameter objects. Each dataset corresponds to one configuration and one controller, and it is filled by its controller. After filling, the input builder will construct the probability density function object from the dataset object through spectrum builders. Their relationships are shown in Figure~\ref{fig:GooStats-input-UML}. The configuration objects are managed using the tree data structure in order to support the parameter synchronization technique, which will be explained in the next section.

\subsection{Parameter synchronization technique\label{sec:parSync}}
\begin{figure}[htbp]
\centering 
\includegraphics[width=0.5\textwidth]{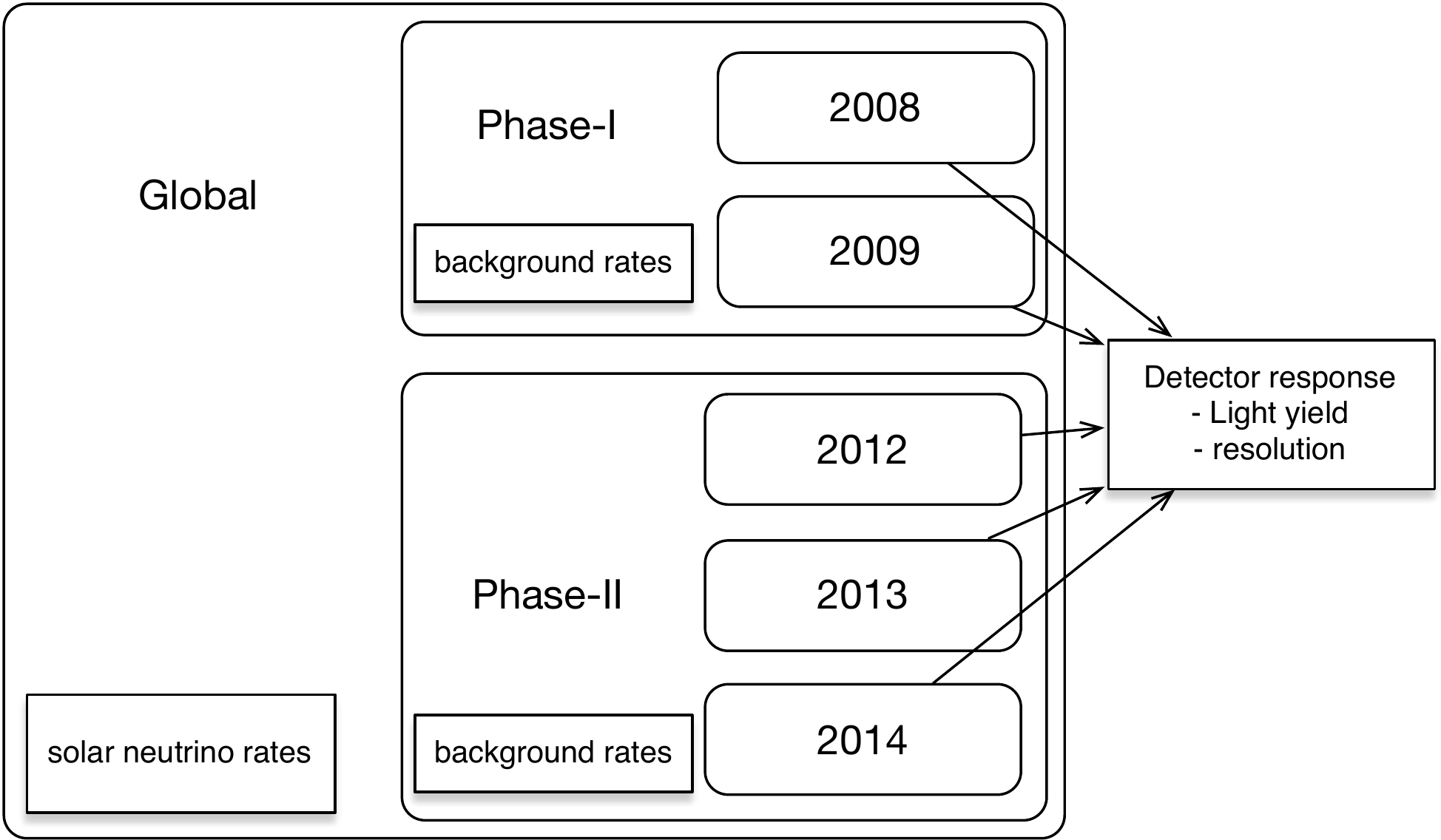}
\includegraphics[width=0.45\textwidth]{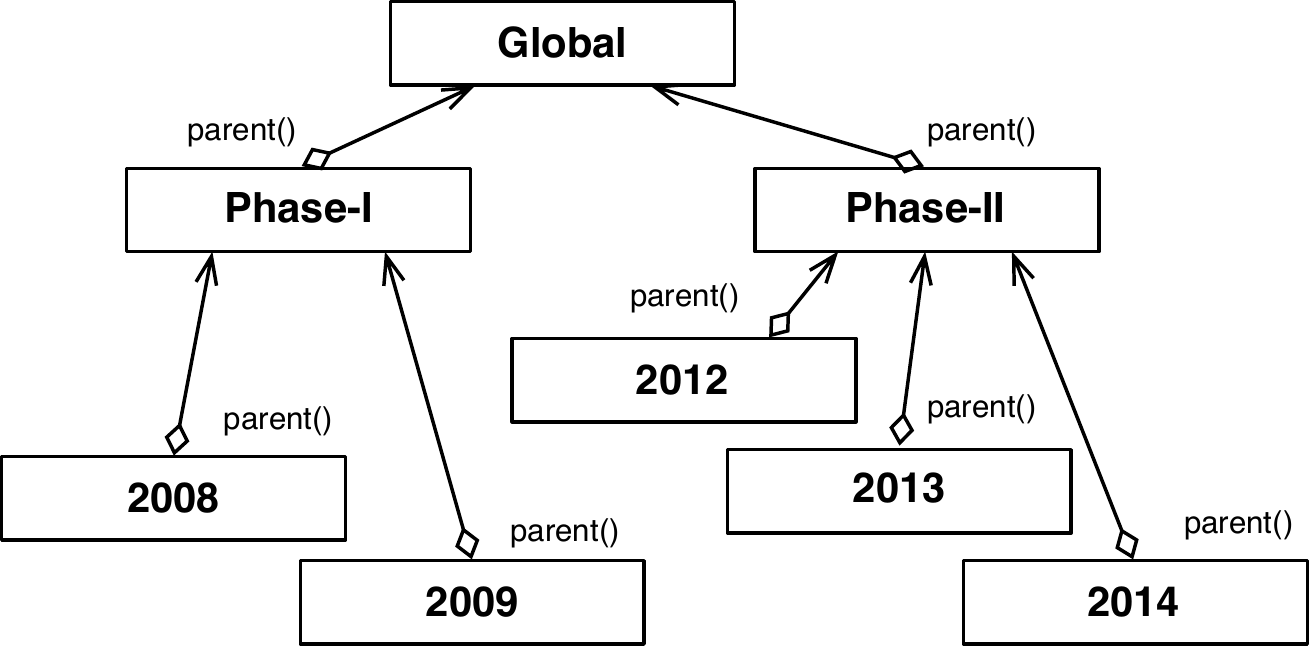}
\caption{\label{fig:parSync} Left: An example synchronization strategy of Borexino Phase-I and Phase-II joint analysis using parameter synchronization technique. Right: The internal data structure. A parameter stored in a parent node will be synchronized among all its children nodes.}
\end{figure}

An analysis technique called parameter synchronization has been developed to improve the model precision. Consider that we can divide the collected data with respect to the volume and time and assign independent detector response parameters for each piece, while requiring different types of fit parameters to be synchronized among datasets at different levels of granularity. For example, Borexino has collected in total around five years of data in period Phase-I and period Phase-II\cite{Bellini2014b,Agostini2017}. With such technique we can divide the collected data into five datasets, each corresponding to one year of data, and fit all datasets simultaneously. The solar neutrino rates should be required to be the same for all datasets because they are stable. The rates of the backgrounds, the radioactive decays, are stable within each period, while the purification campaign between two periods lowered their rates, so it is better to keep their rates the same in each period but independent between two periods. Meanwhile, in Borexino analyses, analytical functions are used to describe the detector response, and because the spectrum is very sensitive to the light yield and the PMTs are dying, it is sensible to keep detector response parameters independent among each year. Such strategy and corresponding data structure are summarized in Figure~\ref{fig:parSync}. 

\begin{figure}[htbp]
\centering 
\includegraphics[width=0.43\textwidth]{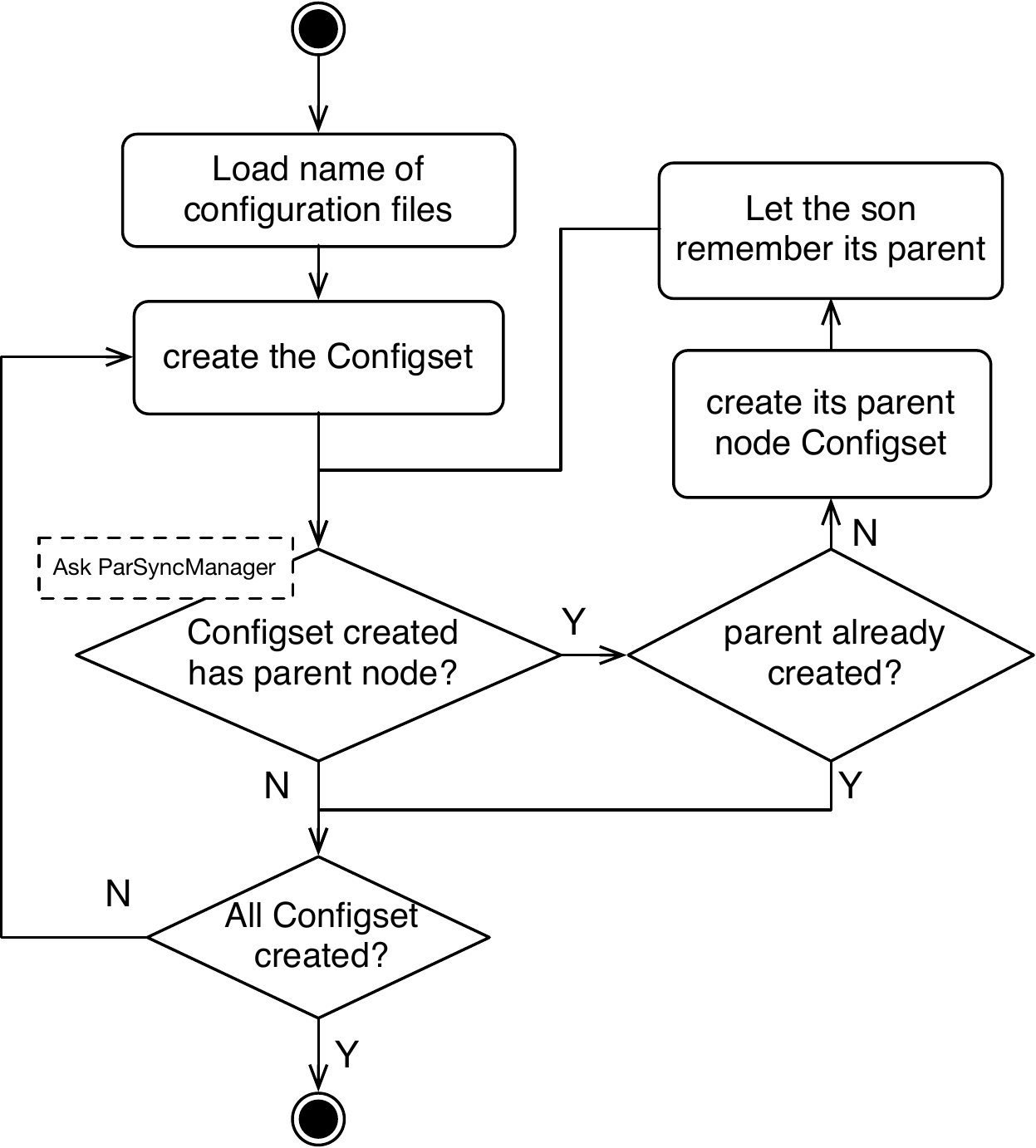}
\hspace{20pt}
\includegraphics[width=0.37\textwidth]{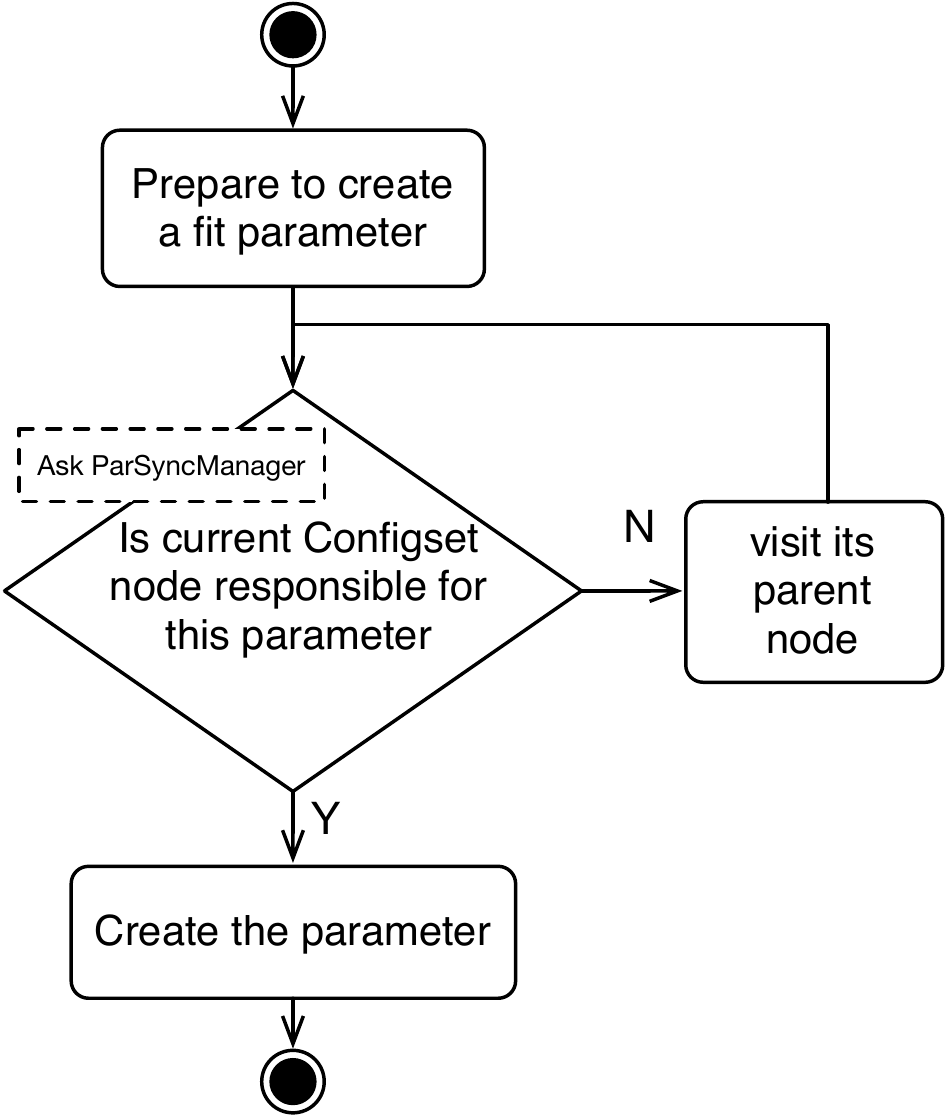}
\caption{\label{fig:CreateParsync} Flow chart of building the tree of configuration objects (left) and creating a fit parameter (right).}
\end{figure}

Such technique is implemented using the decorator design pattern and tree data structure. In \texttt{GooFit}, each fit parameter is represented with an independent object and if the object is shared within various probability density function objects, its value will be synchronized among them. In \texttt{GooStats}, the fit parameter objects are stored in the configuration objects. When filling the fit parameter objects into the dataset object, the synchronization manager will visit the tree until it finds the correct parent node corresponding to the required synchronization level of the parameter. The process of building the tree and that of visiting the tree are summarized in Figure~\ref{fig:CreateParsync}.

\subsection{Detector response functions}
Realistic analytical detector energy response functions inspired by Borexino solar neutrino analysis are built in as standard libraries\cite{Bellini2014b}. Three functions, the modified Gaussian function\cite{Saldanha2012}, the generalized gamma function\cite{Smirnov2008} and the scaled Poisson function\cite{Bohm2014} are included.
%
%

\subsection{Multivariate analysis}
Multivariate analysis can be performed such that information of a secondary observable can be included by adding extra pieces of likelihoods to the likelihood of the main observable\cite{Davini2012}. 
%
%
This method does not require the knowledge of the high-dimensional joint probability density function of considered observables and is faster due to less number of bins. The corresponding probability density function class is \texttt{MultiVariatePdf}.

\subsection{Modification to \texttt{GooFit}}
The minimization engine \texttt{GooFit} is a well-maintained open source project, yet modifications to it have been done to speed up the fitting procedure as well as allowing fast evaluation of the likelihood for statistical analysis.

First, nested kernel calls are avoided and it improved the speed by an order of magnitude. In nVidia GPU, computation tasks are completed in CUDA kernels. 
To evaluate the visible energy spectrum on GPU, hundreds of CUDA kernels, each responsible for one visible energy point, are launched in parallel on GPU execution units. In the original version of \texttt{GooFit}, each CUDA kernel will launch in parallel hundreds of nested CUDA kernels to evaluate the value of the response function on the physics energy mesh to do the detector energy resolution convolution. These nested CUDA kernels make \texttt{CUDA} compiling and linking extremely slow, saturate the warp schedulers during runtime and block the data flow. In the modified \texttt{GooFit} shipped with \texttt{GooStats}, 
the value of response function on two-dimensional visible-energy--physics-energy mesh are evaluated and cached first, and only when all the corresponding CUDA kernels have finished would new CUDA kernels doing detector response smearing convolution be launched. In this way nested calls are avoided, and the speed is improved by an order of magnitude.

Second, evaluated function values are all cached to save computing time. A global \texttt{map} object is added to the \texttt{GooFit} to store the values. When all the parameters of a probability function do not change, there is no need to re-evaluate its values, and caching values saves much time.

\section{Validation and benchmark}
\label{sec:benchmark}
When developing the module for Borexino Phase-II analysis\cite{Agostini2017}, the \texttt{GooStats} software framework was validated against the existing analysis tool. Tests were also designed to measure the overhead. In this section I will present the validation and benchmark results.

\subsection{Validation}
\begin{figure}[htbp]
\centering 
\includegraphics[width=.6\textwidth]{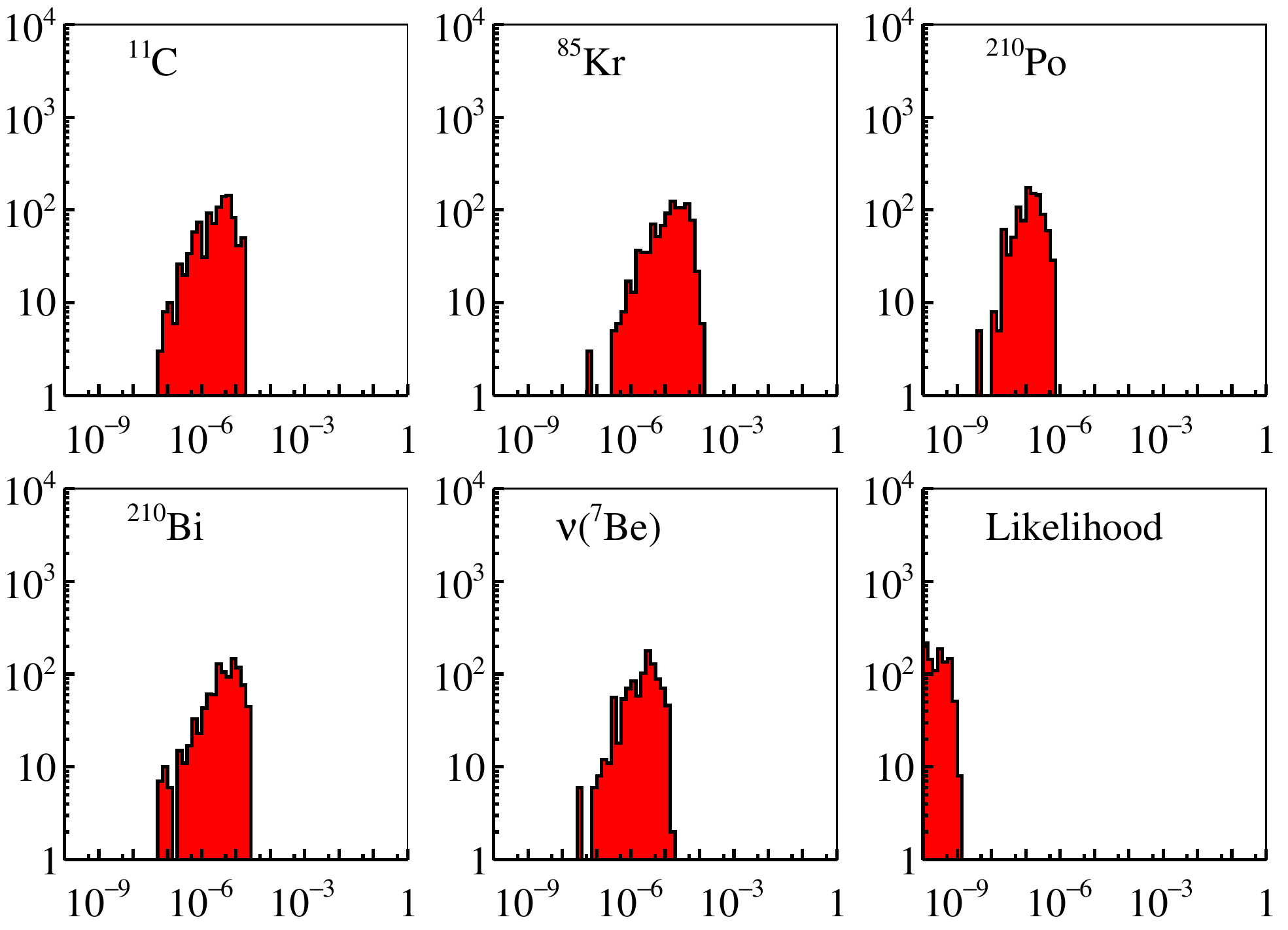}
\caption{\label{fig:validation} The distribution of relative difference \(\left|\Delta_R/R\right|\) of the optimized value of various species rates and the likelihood between the \texttt{GooStats} Borexino module and the existing tool used by the Borexino collaboration.}
\end{figure}

Validation processes have been designed such that the relative difference of the expected number of events between the \texttt{GooStats} Borexino module and the existing software used by the Borexino collaboration should be within \(10^{-12}\) for all bins and all species, and the relative difference of the optimized value of the rates and the likelihood should be smaller than \(10^{-4}\). The comparison results are shown in Figure~\ref{fig:validation}. As can be seen that the results are within required precision. The validation is implemented using the \texttt{GoogleTest} library\footnote{See \url{https://github.com/google/googletest}}.

\subsection{Performance}
The computing time needed to solve a statistical problem depends strongly on the specific application. GooStats can handle out-of-the-box three kinds of fits:
\begin{enumerate}
\item Fitting with the Monte Carlo detector response model, hereafter as "Monte Carlo fit";
\item Fitting with the analytical detector response function model, hereafter as "Analytical fit";
\item Fitting with multiple observables using multivariate likelihood, hereafter as "Multivariate fit".
\end{enumerate}

The time spent on \texttt{GooStats} are mainly used on the parallel computing tasks executed on GPU. The fitting time almost linearly scales with the size of the problem, while the size of the problem is determined by different factors among different types of jobs:
\begin{itemize}
\item For the Monte Carlo fit, the computation heavy part is the evaluating of likelihood on each bin, and thus the size of the problem is the number of bins:
\begin{align}
T = T_0 + N_{E_{\rm vis}} \cdot k 
\end{align}
where $T$ is the total wall time, $T_0$ is the overhead used on loading and parsing input and producing figure outputs etc., $N_{E_{\rm vis}}$ is the number of bins of the visible energy spectrum, $k$ is the speed of the program.
\item For the analytical fit, each expected spectrum is the convolution between the energy spectrum and  the detector response, and thus the size can be defined as the sum of product of the mesh size of the physical energy spectrum and that of the visible energy spectrum:
\begin{align}
T = T_0 +  k_1 \cdot \sum_{i = 1}^{N_{\rm comp}} N^i_{E_{\rm vis}} \cdot N^i_{E_{\rm true}}
\end{align}
where $N_{\rm comp}$ is the number of components, $N^i_{E_{\rm kin}}$ is the number of bins of the physical energy spectrum.
\item For the multivariate fit, much more time is spent on the calculation of the multivariate-likelihoods. The size of the problem can be defined as the following:
\begin{align}
T = T_0 + k_2 \cdot \sum_i M_{i}
\end{align}
where $M_{i}$ is the number of energy slices of secondary observable spectrum.
\end{itemize}

\begin{figure}[htbp]
\centering 
\includegraphics[width=.33\textwidth]{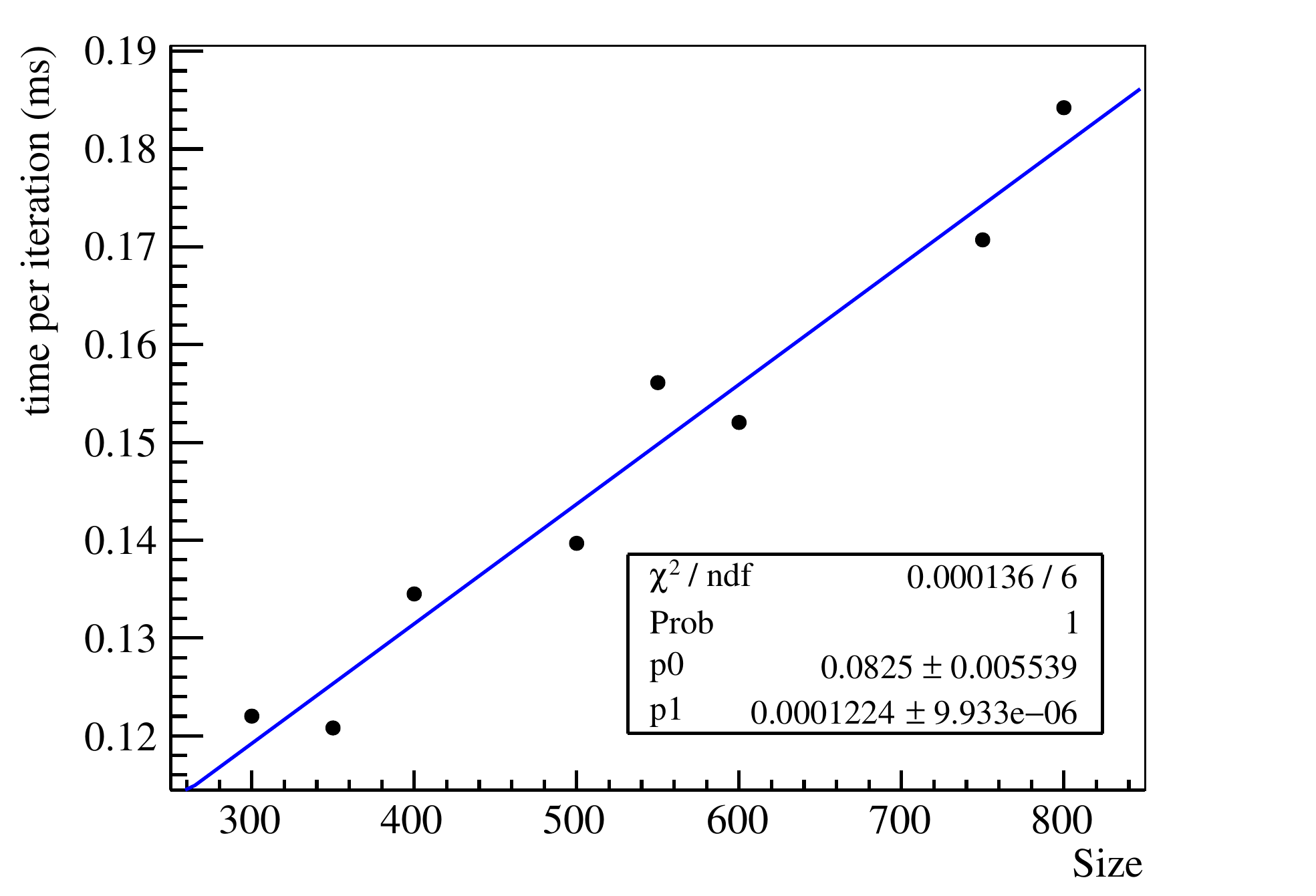}
\includegraphics[width=.32\textwidth]{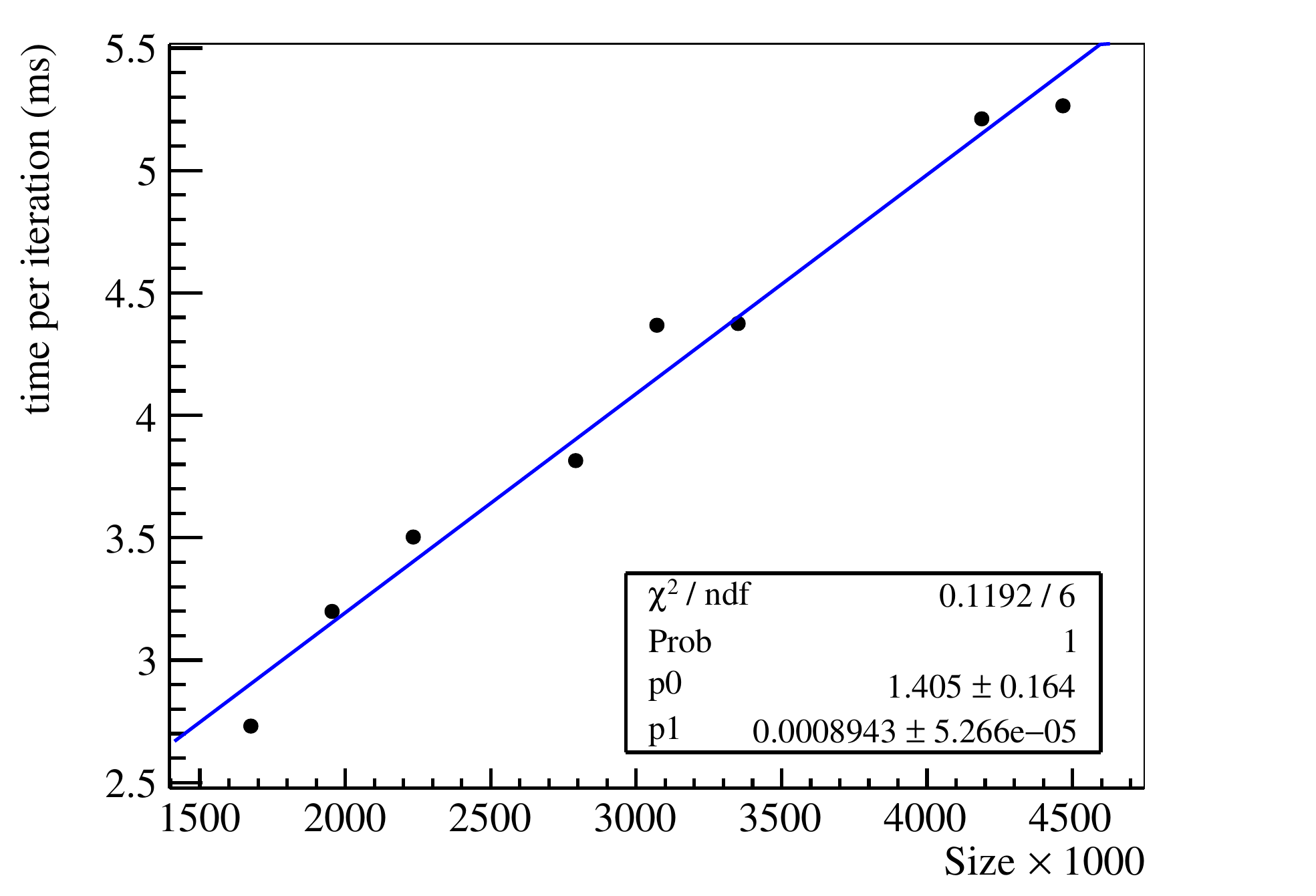}
\includegraphics[width=.33\textwidth]{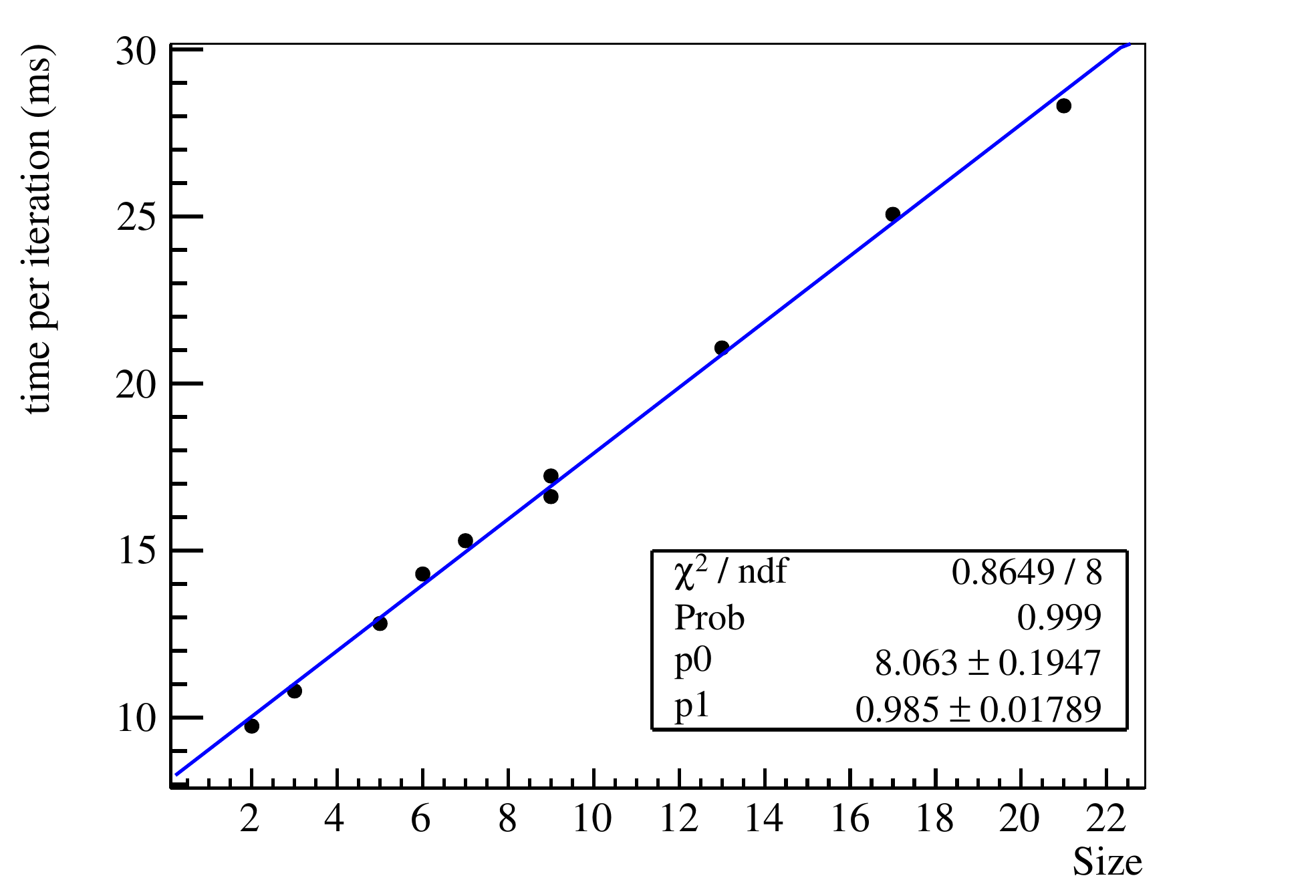}
\caption{\label{fig:benchmark} The time used per iteration during minimization for three types of jobs. Left: Monte Carlo fit. Middle: Analytical fit. Right: Multivariate fit. Least square fits with linear function are also shown.}
\end{figure}

We performed tests to evaluate the overhead. Fitting tasks of different sizes are performed on an nVidia Tesla K20m GPU\cite{NVidiaInc.2012}, whose double precision processing power is 1.18 TFLOPS (tera floating point operations per second). Sizes of fitting tasks are varied by changing the bin size and the fit range. The results are shown in figure~\ref{fig:benchmark}. We can see that the speed of the software on the K20m GPU is satisfactory that the fit time per iteration for a typical-size fitting task in Borexino Phase-II analysis\cite{Agostini2017} ranges from sub-milli-seconds to tens of milli-seconds, depending on the type of the task. The fraction of the overhead, that is, the time spent on parsing configuration files, communicating between CPU and GPU, I/Os, etc., in the total wall time is around 50\% and is also satisfactory.

We also compared the fitting time between \texttt{GooStats} Borexino module and the original software used by the Borexino collaboration for the "Analytical fit" type of job. The results are summarized in Table~\ref{tab:performance}. The speed up is more than 100.

\begin{table}[h]
\caption{Comparison of fitting time between \texttt{GooStats} and original software used by the Borexino collaboration. \(T_{\rm tot}\): total execution time. \(N\): the number of iterations taken to converge in \texttt{MINUIT}. \(T_{\rm it}\): average execution time per iteration. Speed up: \(T_{\rm it}({\rm CPU})/T_{\rm it}({\rm GPU})\).}
\label{tab:performance}
\centering
\begin{tabular}{|c|ccc|ccc|c|}
\hline
& \multicolumn{3}{c|}{CPU} & \multicolumn{3}{c|}{GPU} & \\
Type & \multicolumn{3}{c|}{AMD Opteron(TM) Processor 6238} & \multicolumn{3}{c|}{nVidia Tesla K20m} & \\ 
Size & \(T_{\rm tot}\) (s) & \(N\) & \(T_{\rm it}\) (ms) & \(T_{\rm tot}\) (s) & \(N\) & \(T_{\rm it}\) (ms) & speed up \\
\hline
400 & 27.6 & 1128 & 24.4 & 0.181 & 1346 & 0.135 & 181 \\
350 & 29.4 & 1331 & 22.1 & 0.156 & 1294 & 0.121 & 183 \\
300 & 22.5 & 1239 & 18.2 & 0.243 & 1995 & 0.122 & 149 \\
\hline
\end{tabular}
\end{table}
%

\section{Application examples}
In this section I will present the applications of \texttt{GooStats} to fitting and statistical analyses. The codes are available in the \texttt{github} repository\footnote{See \url{https://github.com/GooStats/GooStats}}. First two subsections are two pedagogical examples and the subsequent two subsections are two realistic problems: the solar neutrino multivariate fit analysis and the medium baseline vacuum oscillation based neutrino mass ordering determination analysis. To fit solar neutrino spectra, user can use built-in libraries and only need to modify the configuration files. For the neutrino mass ordering analysis, user need to add new type of \pdf{} and new type of statistical analysis. The spectra used in this section are all generated using the random number engine.

\subsection{Fit of Gaussian signal plus flat background}
\paragraph{User guide} Consider we would like to extract the rate of a Gaussian signal over a flat background. For this simple task, a \texttt{main} function using the default classes is enough:
\begin{lstlisting}
int main (int argc, char** argv) {
  AnalysisManager *ana = new AnalysisManager();

  InputManager *inputManager = new InputManager(argc,argv);
  inputManager->setInputBuilder(new SimpleInputBuilder());
  GSFitManager *gsFitManager = new GSFitManager();
  OutputManager *outManager = new OutputManager();
  outManager->setOutputBuilder(new SimpleOutputBuilder());
  outManager->setPlotManager(new SimplePlotManager());

  StatModule::setup(inputManager);
  StatModule::setup(gsFitManager);
  StatModule::setup(outManager);

  PrepareData *data = new PrepareData();
  SimpleFit *fit = new SimpleFit();

  ana->registerModule(inputManager);
  ana->registerModule(data);
  ana->registerModule(fit);
  ana->registerModule(outManager);

  ana->init();
  ana->run();
  ana->finish();

  return 0;
}
\end{lstlisting}
An example project is provided in the \texttt{GooStats/Modules/simpleFit} folder. A \texttt{Makefile} is provided for convenience and user can compile this example project with a \texttt{make} command. Users can run the fit with the command 
\begin{lstlisting}
./fit toyMC.cfg out exposure=500
\end{lstlisting}

\begin{figure}[h]
\centering
\includegraphics[width=0.49\textwidth]{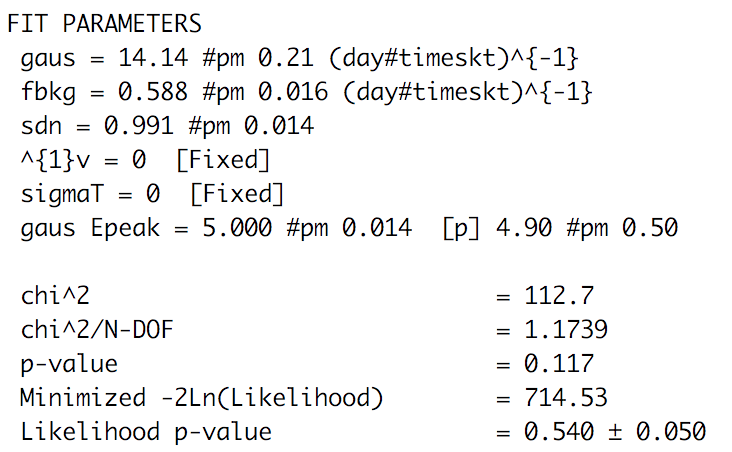}
\includegraphics[width=0.49\textwidth]{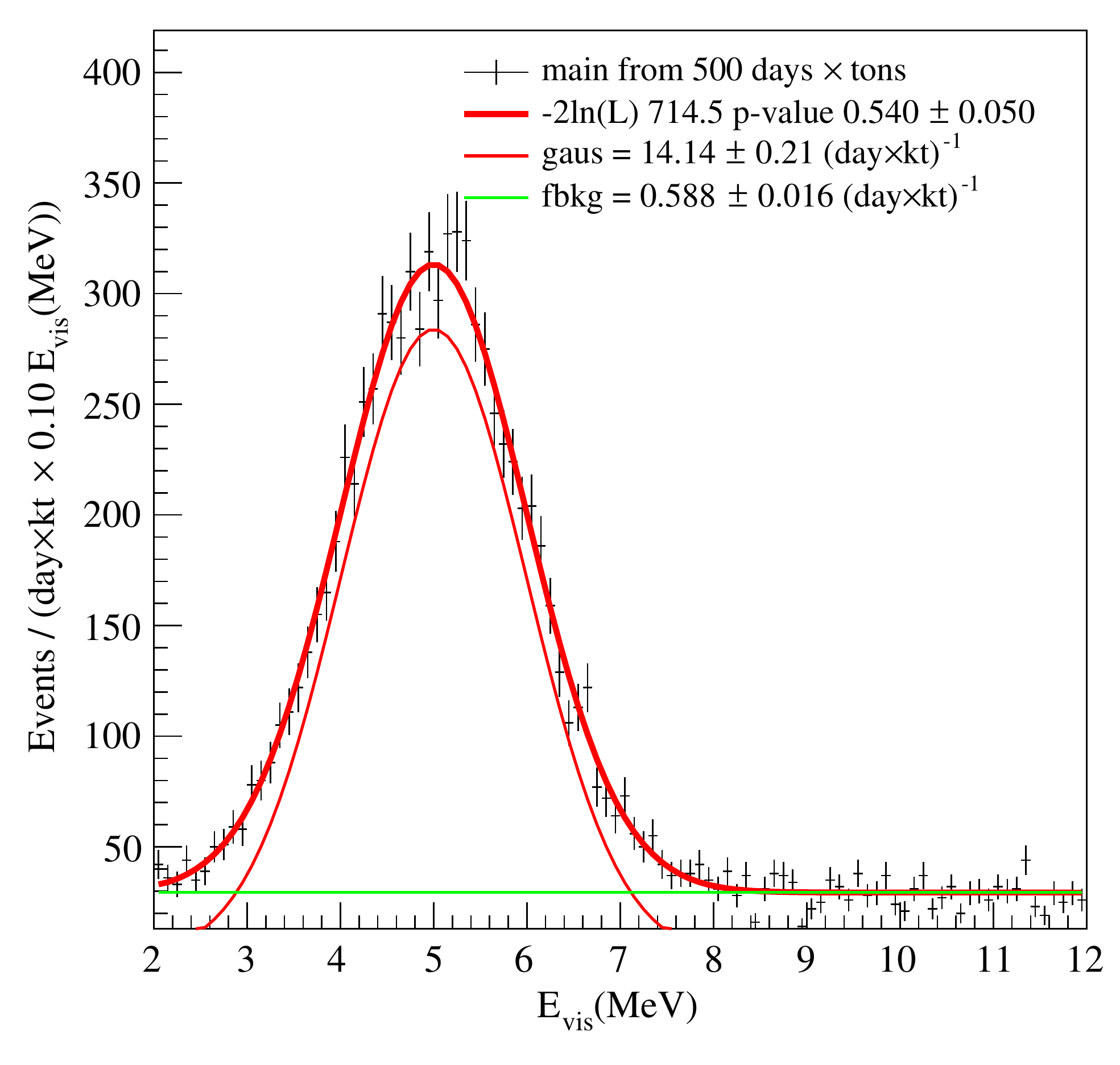}
\caption{Left: screen shot of fit result summary. Right: produced figure in the file format of \texttt{pdf}.}
\label{fig:gausFlat}
\end{figure}

\begin{figure}[h]
\centering
\includegraphics[width=0.61\textwidth]{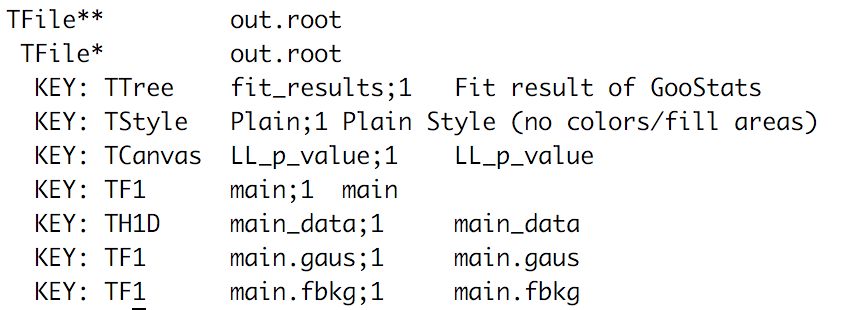}
\includegraphics[width=0.35\textwidth]{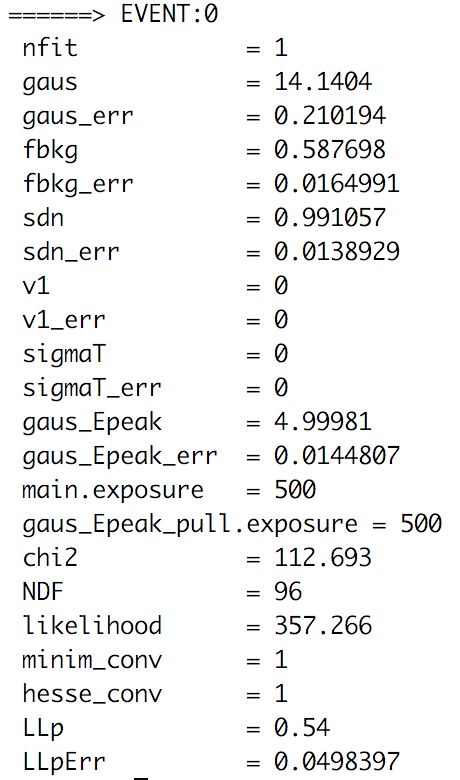}
\caption{Left: content of \texttt{TFile} output. Right: content of \texttt{TTree} output.}
\label{fig:gausFlatTFile}
\end{figure}

\paragraph{Output} At the end, a summary of fit results is printed on the screen. When we would like to see the impact of changing a few fit configurations quickly, this summary would be useful. See Figure~\ref{fig:gausFlat} left. A figure in file format of \texttt{PDF} is also produced, which directly presents how well models describe the data. See Figure~\ref{fig:gausFlat} right. \texttt{GooStats} also produce outputs in \texttt{CERN ROOT} objects. A file in the \texttt{TFile} format containing \texttt{TF1} objects of each components, etc. is produced. See Figure~\ref{fig:gausFlatTFile} left for the full list of objects saved in the file. Sometimes we need the distribution of fit results against many pseudo-experiment spectra, so it would be convenient to have output in the \texttt{TTree} form. It is included in the produced \texttt{TFile}. See Figure~\ref{fig:gausFlatTFile} right.
 
\paragraph{How to set configurations} The contents in the configuration file and on the command line arguments can be understood as the following. In the process of the fit, the binned log-likelihood is minimized:
\begin{align}
\ln\mathcal{L} = \sum_i \ln\left( \dfrac{\lambda^k_i}{k_i!}e^{-\lambda_i} \right) + \sum_j \left(\dfrac{p_j-c_j}{\sigma_j}\right)^2 \label{eq:LL}
\end{align}
where the first term is the Poisson likelihood and the second term is the sum of the pull terms; \(k_i\) is the number of events in the \(i\)-th bin of the visible energy histogram of selected events, \(\lambda_i\) is the expected number of events in that bin, \(p_j\) is \(j\)-th constrained fit parameters, such as the position of the peak, \(c_j\) and \(\sigma_j\) are the centroid and width of the corresponding pull term. The calculation of Equation (\ref{eq:LL}) is performed in the object of \texttt{SumLikelihoodPdf}, and \texttt{GooStats} serves to construct it. In order to construct the \texttt{SumLikelihoodPdf} object, the following inputs are needed:
\begin{itemize}
\item The histogram of the observable of selected events.
\item The list of components, their \pdf{} types.
\item The initial guesses and ranges of rates and response function parameters
\item The list of pull terms, including the name of the constrained parameter, the centroid and the width of the constraint.
\item The exposure
\end{itemize}
They are loaded from the configuration file and the command line arguments in the format of key-value pairs. The pairs on the command line arguments will override the contents in the configuration file.

\subsection{Statistical analysis}
Consider that, in the previously mentioned project, we would like to evaluate the significance of the Gaussian signal. To do so, we define the test statistic as the profile likelihood ratio\cite{Cowan2011}
\begin{align}
t = -2\ln\left(\dfrac{\mathcal{L}(\hat\lambda,\,\hat\theta)}{\mathcal{L}(0,\,\hat{\hat{\theta}}(0))}\right)
\end{align}
where the nominator is the maximized likelihood with the Gaussian signal rate free, the denominator is the maximized likelihood with the Gaussian signal rate fixed to zero. Our task is to obtain its distribution under null hypothesis assumption and to obtain its value corresponding to the data histogram.

First we need to add two lines to the main function. The modified part reads like this:
\begin{lstlisting}
  PrepareData *data = new PrepareData();
  SimpleFit *fit = new SimpleFit();
  DiscoveryTest *discovery = new DiscoveryTest();

  ana->registerModule(inputManager);
  ana->registerModule(data);
  ana->registerModule(fit);
  ana->registerModule(discovery);
  ana->registerModule(outManager);
\end{lstlisting}
An example project is provided in the \texttt{GooStats/Modules/statAnalysis} folder. 

After that we run the following command to obtain the test statistic value corresponding to the data histogram: 
\begin{lstlisting}[language=bash]
./stat toyMC.cfg out exposure=500 DiscoveryTest=default.gaus
\end{lstlisting}
 The log-likelihood-ratio can be retrieved in the \texttt{CERN ROOT terminal} using
\begin{lstlisting}
root [0] fit_results->Scan("likelihood[1]-likelihood[0]");
\end{lstlisting}

At last we run the following command to obtain the test statistic distribution under null hypothesis assumption using the Monte Carlo method:
\begin{lstlisting}[language=bash]
./stat toyMC.cfg nullpdf exposure=500 DiscoveryTest=default.gaus fitFakeData=true repeat=100
\end{lstlisting}
 The distribution can be retrieved in the \texttt{CERN ROOT terminal} using
\begin{lstlisting}
root [0] fit_results->Draw("likelihood[1]-likelihood[0]","minim_conv[0]&&minim_conv[1]");
\end{lstlisting}

\subsection{Solar neutrino spectrum fit}
Solar neutrino produced during the proton-proton fusion in the core region of the Sun can be detected on the Earth using large liquid scintillator detectors, such as Borexino\cite{Mosteiro2015}, KamLAND\cite{Gando2015} and the future experiment JUNO\cite{An2016}, etc. The signal is the electron recoil spectrum formed by the elastic scattering between the solar neutrinos and the electrons in the target. Natural radioactive decays and decays of cosmogenic short-lived isotopes are the main backgrounds, and can be reduced by the purification of the liquid scintillator and putting the detector at deep underground or inside mountains.

The \(^{7}\)Be solar neutrino elastic scattering spectrum has the shape of the Heaviside-step function and has a unique sharp shoulder. It is different from background of decays of \(^{210}\)Bi, \(^{85}\)Kr and \(^{210}\)Po. Its rate can be extracted in a shape analysis using, for example, the maximum likelihood estimation method. The \(pp\) solar neutrino spectrum, instead, has similar shapes of the \(^{14}\)C decay and the pileup events, but softer, and can be extracted if the \(^{14}\)C and pile-up rates are measured independently and constrained. 

\begin{figure}[htbp]
\centering 
\includegraphics[width=.7\textwidth]{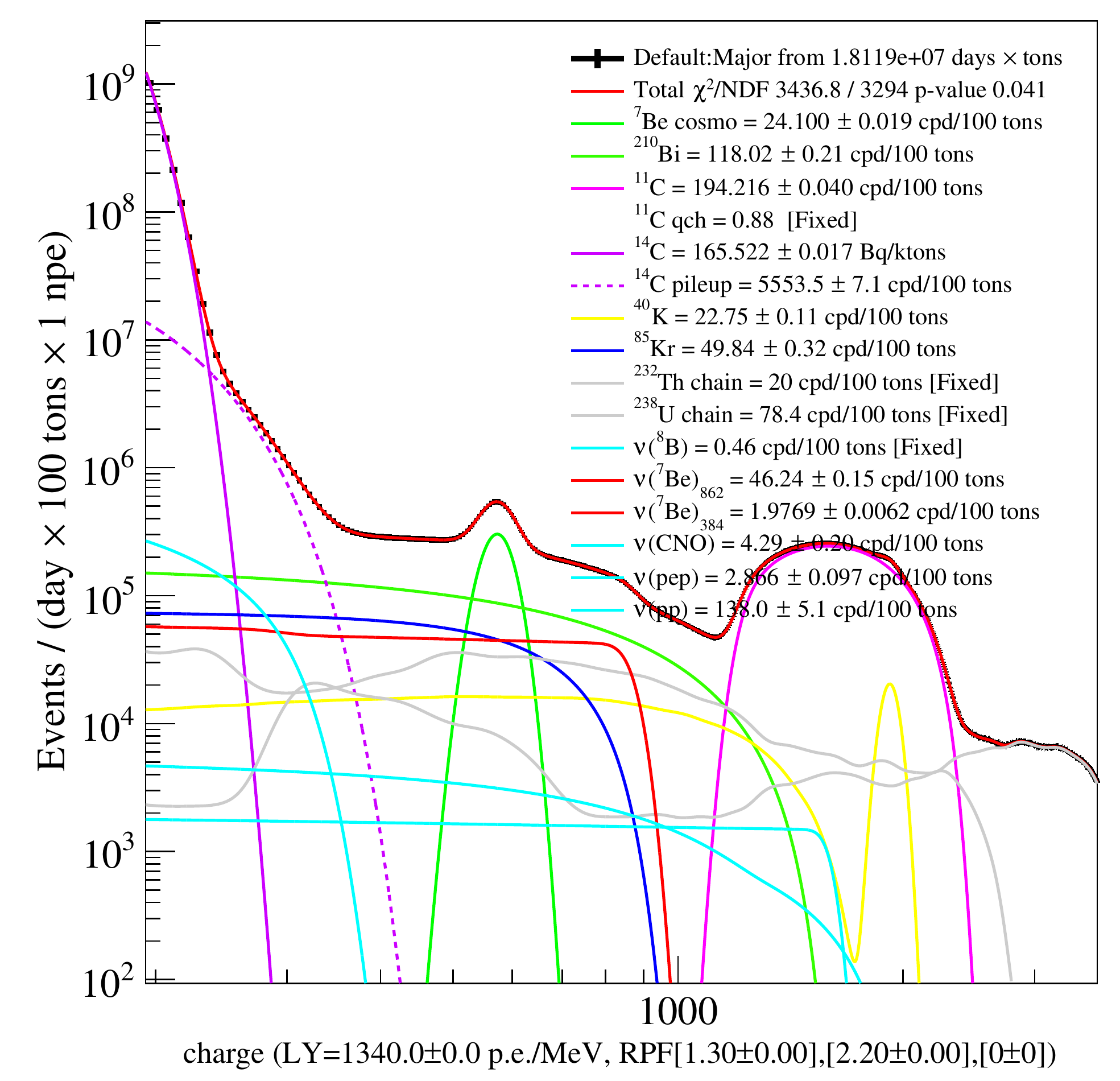}
\caption{\label{fig:solar} An example fit of simulated solar neutrino elastic scattering spectrum in a liquid scintillator detector. Color, line style and line shapes of each species can be easily customized through configuration files.}
\end{figure}

The fit of solar neutrino spectra is similar to the fit of a Gaussian signal plus a flat background. The difference is that in the former the \pdf{}s are more complex. In Borexino, in order to study the systematic uncertainty induced by the inaccuracy of the detector response modeling, the energy smearing is modeled by the convolution with a Gaussian-like response function, and part of the parameters describing the non-linearity and resolutions are left free. There are built-in classes describing such kinds of \pdf{}s, and user only need to specify their types as \texttt{Ana}, \texttt{AnaShift} or \texttt{AnaPeak}. The energy spectra before smearing is constructed by looking for the configuration of the component named as \texttt{xxx\_inner}. For example, to construct \(\nu(^7{\rm Be})\), in the configuration file it should be written that:
\begin{lstlisting}[language=bash]
nu_Be7_type = Ana
nu_Be7_inner_type = MC
nu_Be7_inner_min = 0
nu_Be7_inner_max = 0.68
nu_Be7_inner_nbins = 1000
\end{lstlisting}

An example fit result is shown in Figure~\ref{fig:solar}.

\subsection{Medium baseline neutrino mass ordering determination}

The neutrino mass ordering can be determined in a medium baseline experiment with reactor anti-electron neutrinos, by considering the difference between two neutrino mass ordering on the 3-neutrino vacuum oscillation inference. By fitting the energy spectrum with normal ordering and inverted ordering sequentially then taking the difference of the \(\chi^2\) of two fits, the inverted or normal ordering hypothesis can be rejected at certain confidence level depending on the magnitude of the absolute value of the \(\Delta\chi^2\).

This project is more complex, because built-in classes are not enough to describe the oscillated reactor anti-neutrino spectrum if we would like to leave the neutrino oscillation parameters free. A customized \texttt{SpectrumBuilder} class, which constructs the \pdf{} object, and a customized \texttt{DatasetController} class, which collects needed information, are needed. Besides, three lines need to be modified in the \texttt{main} function:
\begin{lstlisting}
  InputBuilder *builder = new ReactorInputBuilder();
  builder->installSpectrumBuilder(new ReactorSpectrumBuilder());
  inputManager->setInputBuilder(builder);
\end{lstlisting}
An example is given in the \texttt{GooStats/Modules/naive-Reactor} folder. 

\begin{figure}[htbp]
\begin{center}
\includegraphics[width=0.5\textwidth]{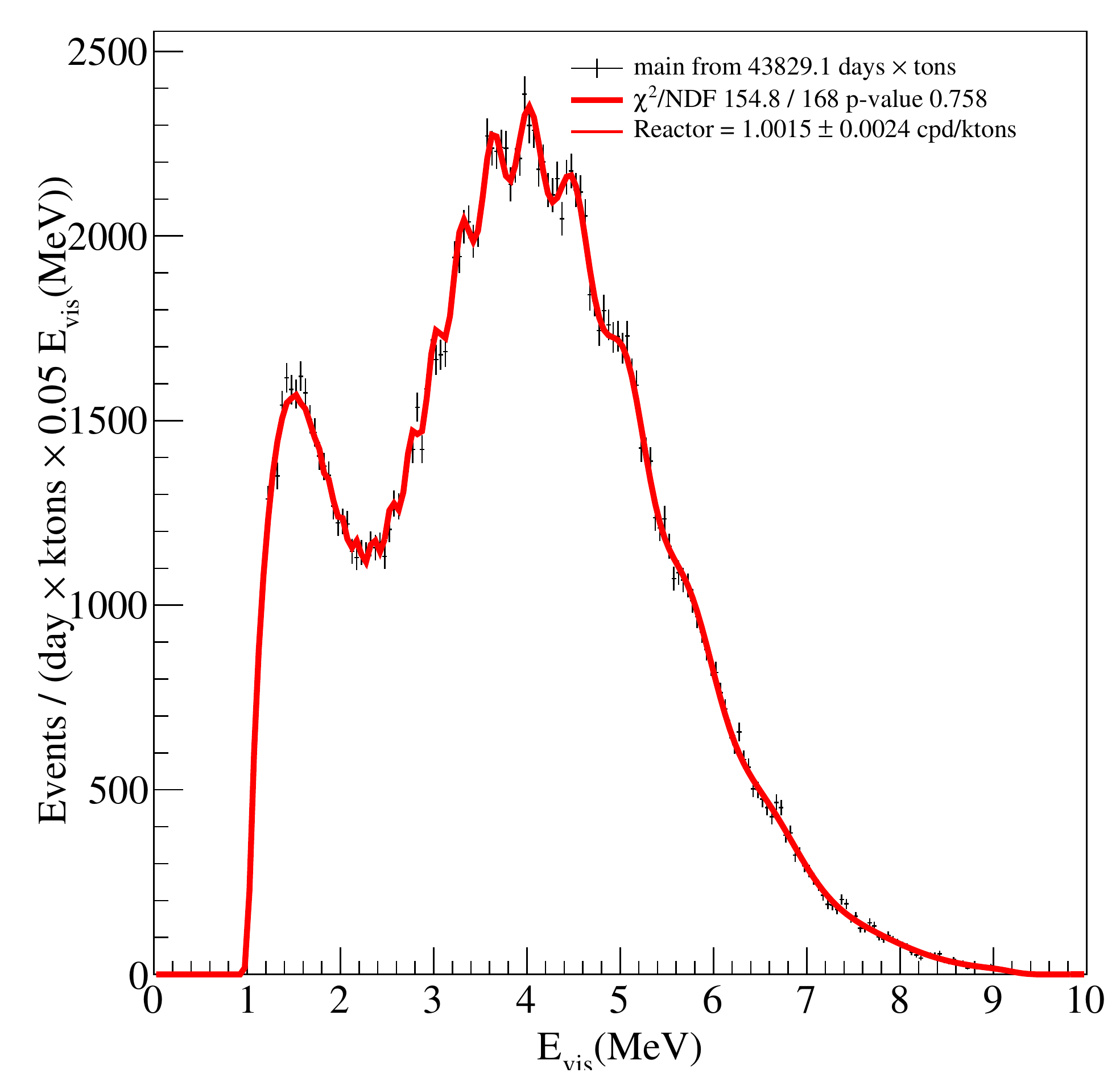}
\end{center}
\caption{Example output of \texttt{naive-Reactor} module.}
\label{fig:output}
\end{figure}

Besides, the statistical analysis also need to be customized. We need to fit assuming normal neutrino mass ordering first, then fit again assuming inverted neutrino mass ordering. To do so, we can write a new \texttt{StatModule} class called \texttt{NMOTest}:
\begin{lstlisting}
bool NMOTest::run(int) {
  if(!GlobalOption()->hasAndYes("fitNMO")) return true;
  auto deltaM2s = getInputManager()->Datasets().front()->get<std::vector<Variable*>>("deltaM2s");
  deltaM2s[1]->value = - deltaM2s[1]->value;
  deltaM2s[1]->lowerlimit = - deltaM2s[1]->upperlimit;
  deltaM2s[1]->upperlimit = - deltaM2s[1]->lowerlimit;
  getGSFitManager()->run(0);
  getOutputManager()->subFit(0);
  deltaM2s[1]->value = - deltaM2s[1]->value;
  deltaM2s[1]->lowerlimit = - deltaM2s[1]->upperlimit;
  deltaM2s[1]->upperlimit = - deltaM2s[1]->lowerlimit;
  getGSFitManager()->run(0);
  getOutputManager()->subFit(0);
  return true;
}
\end{lstlisting}
and register it in the \texttt{main} function:
\begin{lstlisting}
  NMOTest *nmo = new NMOTest();
  ...
  ana->registerModule(nmo);
\end{lstlisting}

An example fit of the normal ordering spectrum with normal ordering hypothesis is shown in Figure~\ref{fig:output}.

\section{Conclusions}
The software framework \texttt{GooStats} has been designed to provide a flexible environment as well as common tools for statistical analysis, especially multivariate spectrum fitting, in particle physics. It is designed to be executed on GPU in order to benefit from the acceleration of parallel computing. The software utilizes the open source minimization engine \texttt{GooFit} and is built upon the \texttt{CERN ROOT} and \texttt{MINUIT} packages. It produces outputs in the format of \texttt{PDF} and \texttt{CERN ROOT} objects. It has been validated against existing tools within \(10^{-13}\) precision, while the fit time is reduced by two orders of magnitude. Benchmark were performed under three typical cases of usage of the software. The performance is satisfactory and there is room for improvement. Example applications including solar neutrino spectrum fit and medium baseline experiment for neutrino mass ordering determination are presented.



\acknowledgments

The author thanks the Borexino collaboration for fruitful feedbacks on the development of the software, thanks Marcin Misiaszek for providing massive GPU computing resources for the validation and benchmark of this software, thanks Matteo Agostini, Le Li and Nicola Rossi for polishing this article, and thanks Eugenio Coccia, Francesco Vissani, and Gran Sasso Science Istitute for providing the initial GPU resources at the early phase of this project.

%
%
%

\bibliographystyle{JHEP}
\bibliography{library}
\end{document}